\documentstyle[12pt]{article}

\begin{document}

\title{Imprint of the Global Hubble Flow on Galactic Rotation
Curves\footnote{UCONN 97-04, November 1997, to appear in the
Proceedings of the Galactic Halos Workshop, UC Santa Cruz, August 1997.}}

\author{\normalsize{Philip D. Mannheim} \\
\normalsize{Department of Physics,
University of Connecticut, Storrs, CT 06269-3046} \\
\normalsize{mannheim@uconnvm.uconn.edu} \\}

\date{}

\maketitle

\begin{abstract}
We identify an apparent imprint of the global Hubble flow on the
systematics of galactic rotation curves, and discuss its possible
implications for gravitational theory.
\end{abstract}

Even though the astrophysics community is currently engaged in an extremely
vigorous debate as to the explicit nature (macho or wimp) of the dark
matter which is widely believed to pervade and even dominate the universe,
it is nonetheless instructive (even if only to sharpen the dark matter
debate itself) to consider the possibility that dark matter may not in fact
actually exist (or at least not in such copious amounts), and that the
apparent need for it actually heralds the explicit failure of the standard
Newton-Einstein gravitational theory on galactic and larger distance scales,
viz. precisely on distance scales which are altogether larger than
the solar system one on which the standard theory was first established.
In order to address this issue it is thus of value to try to identify
phenomenological regularities in the data, regularities which may then
possibly point in the direction of some other candidate gravitational theory,
or which at the very minimum would serve to constrain dark matter dynamics.

Perhaps the most startling aspect of dark matter (other than its very
existence) is that none is actually needed in the solar system at all, and
nor (as noted in various presentations in these proceedings) is any
apparently needed in any sub-galactic system such as a globular cluster or
a molecular cloud, with the standard gravity of the known luminous objects
in such systems providing a completely adequate accounting of their
dynamics.\footnote{Moreover, the very success of standard gravity in its
precision fitting of a binary pulsar, a typical
sub-galactic system, even constrains the density of the nearby, tidally
disrupting, galactic dark matter which is thought to be transiting the Milky
Way with the halo virial velocity. As regards this halo, we note
that while the MACHO project is finding evidence for halo lenses, it is quite
perplexing that their inferred masses would need to be above
the hydrogen burning threshold. Since the luminosity of the LMC is much
larger than that of the resolved source stars being monitored, some detected
events could be due to the
brightening of otherwise unseen LMC source stars as they are lensed,
to then possibly
reduce the true microlensing optical depth.} However, once
one goes to galactic systems not only do the Newton-Einstein gravitational
contributions of the known luminous constituents underestimate the available
data, the associated shortfalls are found to be even bigger on even larger
distance scales such as those associated with systems which contain large
numbers of galaxies.\footnote{In fact, this very need for dark matter
could itself be regarded as being merely a parameterization of the
detected luminous Newtonian shortfalls, with the needed amounts of dark
matter being determined (to be large or small) only
after rather than before the fact.} From this trend it would thus
appear that there is some intrinsic scale associated with the systematics of
luminous Newtonian shortfalls, and thus we shall seek to extract an explicit
one out from galactic data, to find below that not only is there evidence of
such a new scale, but that it intriguingly turns out to be one which is
associated with the global cosmological Hubble flow.

Given the above trend in the growth of the luminous Newtonian shortfall with
distance as we go from galaxies to clusters of galaxies and then beyond, it is
thus suggestive to look at the dependence on (radial) distance of the analogous
shortfalls within individual galaxies themselves. Thus we shall specifically
analyze HI spiral galactic rotation curves, these being the curves which
go out the furthest beyond the optical disk in distance and which precisely
provide the primary evidence for the need for galactic dark matter in the
first place. Now while the flatness of many of these rotation curves is their
immediately most striking and celebrated feature, it is more instructive to
look not at the total observed rotation velocities, but rather to look at the
discrepancies, i.e. to look at the luminous Newtonian shortfalls themselves.
And, indeed, in cases where the total velocities are in fact flat, since the
luminous Newtonian contribution undergoes a Keplerian falloff, it thus follows
that the discrepancies themselves must in fact actually be growing with
distance in such cases. Now out of the currently available set of 33 HI spiral
galaxies K. G. Begeman, A. H. Broeils and R. H. Sanders (MNRAS 249, 523
(1991)) identified a subset of 11 of them as being particularly reliable, a
subset which contains dwarfs (whose rotation curves are typically actually
rising rather than being flat), intermediate spirals (with flat rotation
curves) and bright spirals (with curves which typically slightly decline),
this being a subset which contains galaxies which vary in luminosity by a
factor of more than 1000. Now it was noted by P. D. Mannheim (ApJ 479, 659
(1997); Found. Phys. 26, 1683 (1996)) that even the non-flat rotation curves
of the dwarfs and the bright spirals have discrepancies that likewise grow with
distance out to the largest distances, to thus make growing discrepancies the
most comprehensive qualitative feature of rotation curve systematics. Moreover,
on noting that the centripetal accelerations $(v^2/c^2R)_{last}$ at the last
available data points in each of these 11 galaxies are all found to lie in the
surprisingly narrow range of $1.51\times 10^{-30}$ to $7.25\times 10^{-30}$
cm$^{-1}$ (i.e. surprisingly close given the huge range in luminosity),
Mannheim then found that all of these accelerations could be parameterized
according to the universal three-component relation
$(v^2/c^2R)_{last}=\gamma_0/2+\gamma^{*}N^{*}/2 +\beta^{*}N^{*}/R^2$
where the two new universal constants $\gamma_0$ and
$\gamma^{*}$ take the numerical values $3.06\times 10^{-30}$ cm$^{-1}$ and
$5.42\times 10^{-41}$ cm$^{-1}$ respectively, where
$\beta^{*}=1.48\times 10^5$ cm, and where $N^{*}$ is the total amount
of visible stellar (and gaseous) material in solar mass units in each
galaxy. In addition we now note, that while not being quite as definitive as
the preferred 11 galaxies in the 33 galaxy sample, the other 22 galaxies
(listed for instance in R. H. Sanders, ApJ 473, 117 (1996)) have rotation
curves which in general again all show a trend of increasing discrepancy with
distance. Moreover, for this larger group the centripetal accelerations
at the last detected points are all found to fall in the range
$1.32\times 10^{-30}$ to $5.38\times 10^{-30}$ cm$^{-1}$, i.e. just
the same narrow range found for the other 11 galaxies.

Since there is nothing significant about the last detected data points (they
are fixed by the sensitivity of the 21 cm line detectors and not by any
dynamics within the galaxies themselves), the existence of this empirically
found structure for $(v^2/c^2R)_{last}$ should thus be taken as being
phenomenologically significant, and it should thus serve as a constraint on
all theories of galactic rotation curves. Moreover, since the numerical value
extracted out for the parameter $\gamma_0$ turns out to be of order the inverse
of the Hubble radius, the rotation curves would thus appear to be endowed
with a cosmological imprint. Further, on recognizing that the $\gamma_0/2+
\gamma^{*}N^{*}/2$ term is serving as a linear potential, we thus see that it
its inferred numerical value immediately implies that it would be negligible
for sub-galactic distances where the luminous Newtonian $\beta^{*}N^{*}/R^2$
term would then dominate, while becoming ever more important on galactic and
larger distances scales. The scale associated with this linear potential
thus precisely characterizes when there should or should not be any luminous
Newtonian shortfall.

Now that we have identified this pattern it is immediately natural to look
for a dynamics which might produce it, and indeed Mannheim has noted (as
discussed in his above papers which give detailed related references) that
its emergence is actually quite natural in fourth order conformal gravity, a
fully covariant, pure metric based candidate alternative to standard Einstein
gravity. Indeed, what is found there is that the familiar standard gravity
Schwarzschild metric exterior to a star of Schwarzschild radius $2\beta^{*}$
is generalized in conformal gravity to
$-g_{00}=1/g_{rr}=1-2\beta^{*}/r +\gamma^{*}r$, to thus yield,
for a system of $N^{*}$ stars, the asymptotic
$(v^2/c^2R)_{last}=\gamma^{*}N^{*}/2 +\beta^{*}N^{*}/R^2$ on large enough
distance scales. We thus see that the departure from standard gravity is
described by none other than a linear potential. However, unlike Newtonian
potentials, linear potentials are not asymptotically negligible, and thus,
again unlike Newtonian gravity, it is no longer possible to neglect the
gravitational effects due to matter exterior to any system of interest. Thus
we need to determine what effect the rest of the matter in the universe might
have on motions within individual galaxies. And, quite remarkably, it was
further shown that in conformal gravity the entire Hubble flow is found to
act on such galaxies just like a universal linear potential term (generated by
the scalar curvature $k=-\gamma_0^2/4$ of a necessarily topologically open
cosmology), to give the additional $(v^2/c^2R)_{last}=\gamma_0/2$ term just as
desired. With these two linear potential terms, conformal gravity was then
able to yield parameter free fits to the all of the rotation curve data
points of the selected 11 galaxies (i.e. those at all radial distances and not
just the furthest ones) without any need for dark matter.\footnote{In
these proceedings J. F. Navarro reports on work with C. S. Frenk and S. D. M.
White in which they derived a cosmology based dark matter model with generic
density $\rho(r)=\rho_0/r(1+r/r_s)^2$. It is thus of
interest to note that in the limit of large $r_s$, a galactic halo with such a
density would act precisely the same way as a linear potential. At the present
time their model is not yet detailed enough to determine exactly what specific
luminous matter distribution (i.e. what particular luminosity and optical disk
scale length) would be trapped in any particular halo with specific values for
$\rho_0$ and $r_s$, and so they currently do standard dark matter fitting to
rotation curves by adjusting dark to luminous matter distributions galaxy by
galaxy. It might thus be quite instructive to see whether cosmological dark
matter dynamics could reproduce the universal structure found for
$(v^2/c^2R)_{last}$ with the same facility as conformal gravity.} We thus
identify an imprint of cosmology on galactic rotation curves, and suggest that
it is its neglect which	has generated the conventional appeal to dark matter.
This work has been supported in part by the Department of Energy under grant
No. DE-FG02-92ER40716.00.

\end{document}